\documentclass[twocolumn,secnumarabic,amssymb, 
 aps, prb,superscriptaddress]{revtex4-1}

\def\H{${\cal H}$}
\def\HN{\H-$N$ }
\def\pwz{ f_z}
\def\nwz{ f_z^*}

\def\Tx{\Theta_x}

\newcommand{\be}{\begin{equation}}
\newcommand{\ee}{\end{equation}}

 \usepackage{graphicx}
 \usepackage{epstopdf}

\begin{document}

\title{Dirac loops  in trigonally connected 3D lattices}
\author{Kieran Mullen}
\affiliation{Department of Physics and Astronomy, University of Oklahoma, Norman, Oklahoma 73069, USA}
\author{Bruno Uchoa}
\affiliation{Department of Physics and Astronomy,  University of Oklahoma, Norman, Oklahoma 73069, USA}
\author{Daniel~T.~Glatzhofer}
\affiliation{Department of Chemistry and Biochemistry,  University of Oklahoma, Norman, Oklahoma 73069, USA }
\author{Bin Wang}
\affiliation{School of Chemical, Biological and Materials Engineering,  University of Oklahoma, Norman, Oklahoma 73069, USA }

\begin{abstract}

We consider different generalizations of the honeycomb lattice to three dimensional structures. We address the
family of the hyper-honeycomb lattice, which is made up of alternating layers of 2D honeycomb  nano-ribbons, with each layer rotated by $\pi/2$ with the layer below.  When the orbitals of the lattice sites  are symmetric with respect to the planes of the trigonal links, these structures
 can produce a Dirac loop, a closed line of Dirac nodes in momentum space. For orbitals that break that symmetry, such as the carbon $p$-wave orbitals, hyper-honeycomb lattices do not possess the loop structure. 
 We also consider a new structure, the screw hyper-honeycomb, in which the successive layers of parallel units are rotated by $2\pi/3$.  This structure has a Dirac loop  regardless the symmetry of the onsite orbitals. We discuss the implementation of those systems in optical lattices.\end{abstract}

\maketitle

\section{ Introduction.} 
\label{sec:intro}

Materials such as graphene \cite{GrapheneReview},  3D Dirac and Weyl semimetals \cite{Hasan, Qi} have in common the presence of a Fermi surface that is described by a discrete number of Dirac points.  Another possibility, Dirac line semimetals, are materials where the Fermi surface is described by a line of Dirac nodes \cite{DiracLoopPRL, Kane2, Cava, Weng, Yu, Fu, Volovik, Chen}. One interesting example refers to the case where the Fermi surface is a Dirac loop, a closed line in momentum space where the electrons have a vanishing density of states and about which it varies linearly, as for massless Dirac fermions.  Dirac loops are conceptually interesting due to their quantized conductivities \cite{DiracLoopPRL, Vish, Aji}, and the possibility of a 3D quantum Hall effect \cite{DiracLoopPRL}. 

In honeycomb lattices, the Dirac point is special because it is protected by time reversal and inversion symmetry, and it does not require any spin-orbit interaction or band inversion. Natural extensions of the planar trigonal connectivity of the honeycomb lattice in 3D include the family of harmonic honeycomb lattices, which have been studied both theoretically and experimentally in the context of the Kitaev model \cite{Kitaev, Mandal, Kim, Kimchi}. Here we investigate simple tight binding lattice models with trigonally connected structures and study conditions for the realization of Dirac loops.  We present a systematic study of different families of lattices.  

 In section \ref{sec:hhl}   we examine the harmonic honyecomb lattices, which consist of perpendicularly oriented honeycomb lattice strips.  The Dirac loop is present in tight binding models whenever the on-site orbitals preserve the reflection symmetry of the planes set by the trigonal links, as shown in Fig. 1. This is the case for $s$-wave states in hyper-honeycomb lattices considered in  Ref. [\onlinecite{DiracLoopPRL}]. 
 In the case of tight binding lattices  with higher angular momentum states that break that symmetry, such as $p$-wave orbitals in carbon atoms,  the Dirac loops collapse at the edge of the Brillouin zone and are not observable.  Hence,  Dirac loops cannot be observed in simple carbon realizations of those lattices, but could possibly be realized in other materials and in optical lattice realizations of those structures.  
 
In  section \ref{sec:screw} we
 look at trigonal lattices that have different screw symmetries than the hyper-honeycomb lattices.  We address the family of screw hyper-honeycomb lattices, where the planes of the ribbons are rotated along the screw axis by $2\pi/3$ rather than at right angles. These lattices display a Dirac loop, regardless the symmetry of the states in the lattice sites. 
 
 For $s$-wave states,  those lattices also have Fermi surface points at the corners of the Brillouin zone (BZ), which has hexagonal symmetry.  The energy dispersion around those points is parabolic in plane of the loop and linear in the perpendicular direction. The crossing point has a two-fold band degeneracy per spin, has chiral quasiparticles but zero Chern number. For  higher angular momentum states, those points split into small Dirac loops. While screw hyper-honeycomb lattices can be challenging to realize in solid state structures due to chemical stability, they may have promising realizations in other physical systems. We conclude with a brief discussion about possible implementations of those families of structures in optical lattices.

\begin{figure}[b]
 \centering
 \vspace{0cm}

\includegraphics[width= 3.3in]{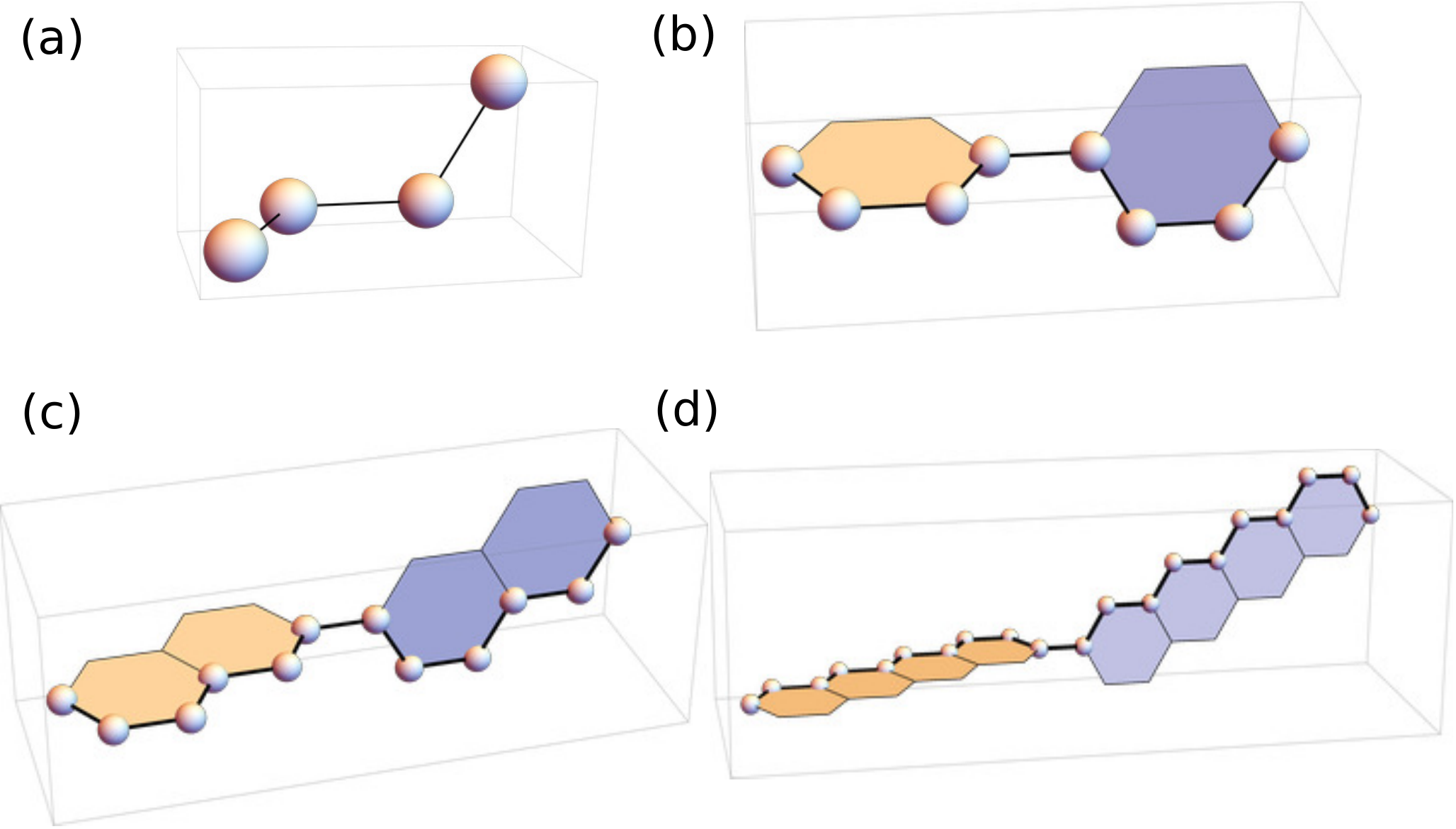}

  \caption{{\small (Color online) Unit cells in the family of the hyper-honeycomb lattice.  All create simple lattice structures where all atoms are connected by three co-planar bonds spaced by $120^\circ$, and are drawn with the positive z-axis pointing to the right, along the screw axis. a) The hyper-honeycomb lattice ($\mathcal{H}$-0), with a four atom unit cell. The full structure consists of a lattice  parallel sets of vertical and horizontal
  linear chains connected by links in the z-direction. b) An eight atom unit cell ($\mathcal{H}$-1).   The horizontal and vertical chains are expanded to strips of single hexagons of graphene.   c)  The
  \H-$2$ unit cell, with 12 atoms,  and  d) the \H-$4$ unit cell, with 20 atoms.  }} 
 \label{fig:hyperhoneycomb}
 \end{figure}

\section{Hyper-Honeycomb Lattice Family} 
\label{sec:hhl}

Our discussion starts with the simplest family of structures shown in Fig.\ref{fig:hyperhoneycomb}a, known as the harmonic honeycomb lattices, $\mathcal{H}$-$N$. Those structures belong to the family of the hyper-honeycomb lattices,  $\mathcal{H}$-0, shown in Fig. 1.  Each structure  consists of a lattice of perpendicular honeycomb nano-ribbons $N$ hexagon units wide,  containing a screw symmetry and a glide symmetry.  Each nano-ribbon is rotated in $\pi/2$ around the screw axis. The result is a structure with planar trigonally connected sites but in 3D.

 The tight binding basis  is a function of  the form $\psi_{\alpha,\mathbf{k}}(\mathbf{r}) = \phi_\alpha(\mathbf{k}) \, e^{i \mathbf{k} \cdot \mathbf{r}}$,  with  $\alpha=1,2,3,\ldots ,N$ labeling  the components of a $N$ vector $\Phi_\mathbf{k}$, which describes the amplitudes of the electronic wavefunction  on
the $N$ atoms in the unit cell. The tight binding Hamiltonian satisfies the eigenvalue equation 
$
 { \mathcal{H}}  \Phi_\mathbf{k} = E  \Phi_{\mathbf{k}}
\label{eq:Htb}
$
where  
\begin{equation}
{\cal H}_{\alpha,\beta}=  \sum_{\vec \delta_{\alpha,\beta}} t_{\alpha,\beta}  e^{i \mathbf{k}\cdot \vec\delta_{\alpha,\beta}}
\end{equation}
and $t_{\alpha,\beta}$ is the hopping energy between nearest neighbors sites separated by
the vector $\vec \delta_{\alpha ,\beta}$ connecting  an atom of the kind $\alpha$ with its nearest neighbor of the kind $\beta$.  The sum is carried over all nearest neighbor vectors $\vec\delta_{\alpha,\beta}$ among any two given species of sites, $\alpha$ and $\beta$. In explicit form, the \H-0 Hamiltonian is:
\be
\mathcal{H}^{(0)}_{\alpha\beta}=t\left(\begin{array}{cccc}
0 & \Theta_{x} & 0 &\gamma  \nwz \\
\Theta_{x}^{*} & 0 &\gamma \pwz & 0\\
0 &\gamma \nwz & 0 & \Theta_{y}    \\
\gamma \pwz   & 0 & \Theta_{y}^{*} & 0\end{array}\right)
 \label{hmat}
\ee
where  $f_z= e^{ik_za}$,  
\begin{equation}
\Theta_{i}=2\mbox{e}^{ik_{z}a/2}\cos(\sqrt{3}k_{i}a/2)
\end{equation}    
with $i=x,y$  and $a$ is the lattice constant.  While all the bonds to a given site atom are co-planar and trigonal,  the bonds of {\it adjacent}  atoms need not lie in the same plane.   The variable $t$ is the hopping parameter for sites in the same ribbon and $t\gamma$ is the hopping for NN sites between two adjacent ribbons rotated in $\pi/2$. 

At the tight binding level,  the orbital sites can be approximated by hydrogen atom wave functions, whose angular part is described in terms of spherical harmonics.  One can account for the effect of the rotation along the screw axis through the overlap hopping integral.  For  $s$-wave states,  the rotation has no effect in the orbital overlap between sites at the edge of the ribbons ($\gamma=1$). For  $p$-wave states,  such as the case of carbon atoms, $\gamma \sim \cos^2 \varphi$ in leading approximation in the tight binding potentials, where $\varphi$ is the rotation angle. Therefore, for screw axis rotations of $\varphi=\pi/2$,  $\gamma=0$. The effects of this rotation for $p$-wave orbitals is addressed below.  The variable $\gamma$ is non-zero by symmetry for higher angular momentum states that preserve the reflection symmetry with respect to the planes of the trigonal links at the edge between two strips (see Fig. 1).

The structure of  matrix (\ref{hmat}) is preserved for all \HN lattices:
\be
\mathcal{H}^{(N)}_{\alpha\beta}=t\left(\begin{array}{cccccccccc}
0 & \Theta_{x} & 0 & \dots & & &  &\dots  &  0& \gamma \nwz\\
\Theta_{x}^{*} & 0 &\pwz & 0 & \dots &  & & &   & 0 \\
 0 & \nwz&0 & \Tx  & \ddots & & &  &  & \vdots  \\
 \vdots  & & \Tx^* &0 & \pwz & & &  &  & \\
 & & &  \nwz &0 & \ddots  & & &   & \\
  & & & & \ddots& \ddots  &  \ddots & \ddots  & & \\
 & & & & &\nwz & 0 &  \Theta_y & 0& \\

\vdots & & & &  & &\Theta_y^* & 0&   \pwz & 0 \\
0&  &  & & &  & 0 & \nwz  & 0 & \Theta_{y}\\
\gamma \pwz &0 & \dots &  &  & &  &  0 & \Theta_{y}^{*} & 0\end{array}\right)
 \label{hmatN}
\ee
producing a banded matrix with elements in the off-diagonal corners.  The upper left half diagonal elements involve $\Theta_x$, while the lower right half
diagonal elements contain
$\Theta_y$;  The four skew-diagonal elements still have a factor of $\gamma$.

 \begin{figure}
 \centering
 \vspace{0cm}
\includegraphics[width=3.3in]{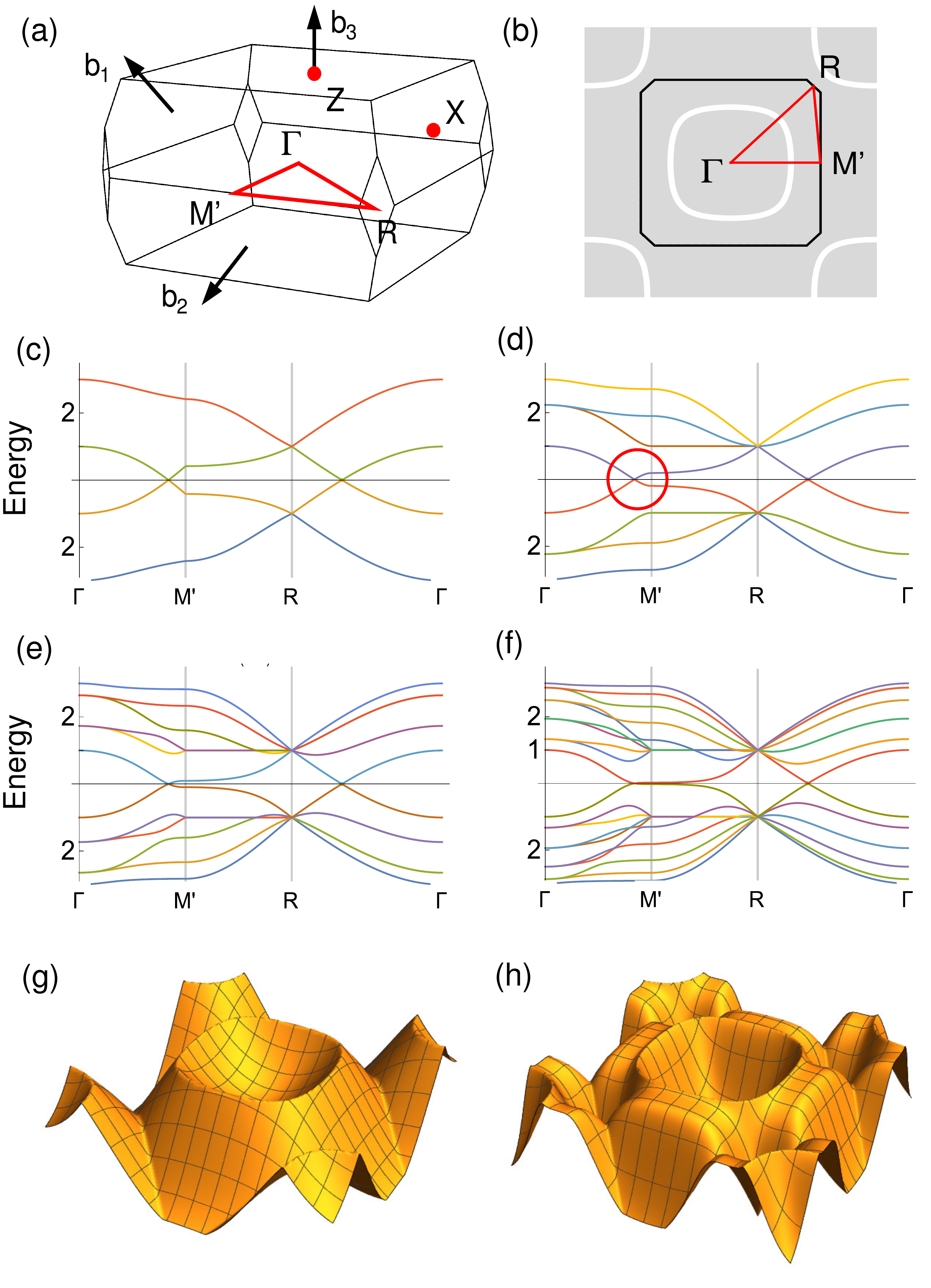} \vspace{-0.5cm}
  \caption{{\protect\small (Color online) a) 3D Brillouin zone of the \H-$0$ hyper-honeycomb lattice. b) BZ in the  $k_z=0$ plane showing the Dirac loop lines (solid white). Black line: boundary of the BZ, centered at the $\Gamma$ point.    Energy spectra of Eq. (2) for $s$-wave orbitals ($\gamma=1$)  in units of the hopping $t$ plotted  along the path shown in the red line of panels a) and b)  for the (c)  $N=0$, (d) $N=1$, (e) $N=2$ and (f) $N=4$ cases.   The low energy bands cross along the Dirac loop present in all structures.  The optical gap outside the loop around decreases rapidly around the $M^\prime$ point (red circle) as $N$ increases.  Energy surfaces at the $k_z=0$ plane for  (g) $\mathcal{H}$-$0$ and (h)   $\mathcal{H}$-$4$ lattices.  }} 
 \label{fig:hhl}
 \end{figure} 
 
  The reciprocal lattice  for the \H-0 lattice is generated by the vectors $\mathbf{b}_1=(2\pi/\sqrt{3}a,0,\pi/3a)$, $\mathbf{b}_2=(0,2\pi/\sqrt{3}a,-\pi/3a)$ and $\mathbf{b}_3=(0,0,2\pi/3a)$, as shown in Fig. 2a, and has four high symmetry points, $\Gamma,\,R,\,X,$ and $Z$.   The BZ of the higher $\mathcal{H}$-$N$ structures is quite similar, with the exception of  the $\mathcal{H}$-1 structure, which is tetragonal and has  generators making 90$^\circ$ angles with each other, 
$ \mathbf{b}_1 = (2\pi/\sqrt{3},0,0)$, 
  $\mathbf{b}_2 = (0,2\pi/\sqrt{3},0)$, and
    $\mathbf{b}_2 = (0, 0, \pi/3,0)$. 
All other structures including the $\mathcal{H}$-0 are triclinic.   

\begin{figure}[t]
 \centering
\includegraphics[width=3.2in]{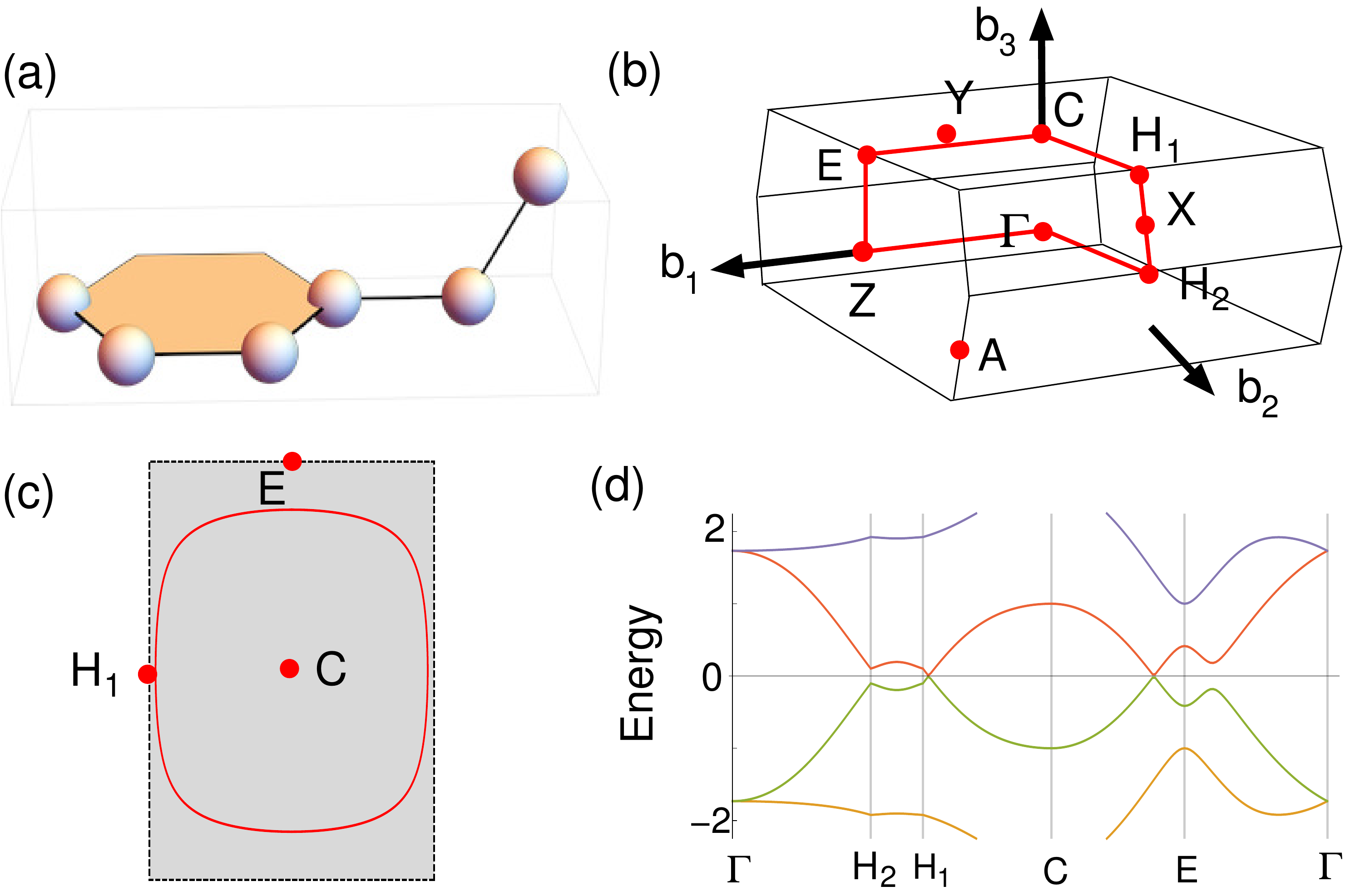}
  \caption{{\protect\small (Color online)  Analysis of the asymmetric (1,0) structure. 
  (a)  The unit cell for the $(1,0)$ structure, which consists of parallel horizontal honeycomb nanoribbons at right angles to layers of parallel, vertical
  chains.  (b)  The Brillouin zone for this structure, which lacks the rotational and screw symmetry of the $\mathcal{H}$-$N$ structures.     (c) Slice at the flattened corner of the BZ for $\gamma=1$.   The Dirac loop (red line) is  centered around the $C$ point, at the edge of the zone.  (d)  Slices of the energy bands along the red path shown in (b).   }} 
 \label{fig:asymm}
 \end{figure}

The harmonic lattices $\mathcal{H}$-$N$ are only a  symmetric subset in a hierarchy of possible lattices that can be  made by perpendicular chains of trigonally connected  atoms.   In Fig. 3a  we show a typical asymmetric  unit cell.  When repeated it
creates a structure in which  the horizontal units are single hexagons as in the \H-$1$ structures, while the vertical  units are simple chains as in the \H-$0$ structure. This asymmetric unit cell in turn breaks screw symmetry and  produces an asymmetric Brillouin zone, with monoclinic symmetry, as shown in Fig.\ref{fig:asymm}b.  Its BZ has 16 high symmetry points, including   $\Gamma$, $H_1$, $H_2$, $C$,  $E$, $Z$, $A$, $X$ and $Y$ points. The generators in the reciprocal space are the vectors $\mathbf{b}_1 = (2\pi/3,0,0))$, $\mathbf{b}_2 = ((0,2\pi/\sqrt{3},-2\pi/9)$, and $\mathbf{b}_3 = (0,0,4\pi/9)$. 
We refer to the structure in Fig. 3a as the (1,0) lattice.  In general, we define the  $(n,m)$ structures as formed by two subsets of alternating honeycomb strips rotated by 90$^\circ$,    with the integers $n_x,\,n_y$ indexing their respective widths in number of honeycomb hexagons along the $x$ and $y$ directions ($n=0$ means a zigzag chain of atoms).

\subsection{s-wave states}

We first consider the case of $s$-wave  orbitals, where $\gamma=1$.  In this case all $\mathcal{H}$-$N$ lattice
  Hamiltonians have a zero energy eigenvalue  along the curve defined by $k_z=0$ and
\be
4 \cos{ \left( { \sqrt{3}} k_x a/2\right) }  \cos{\left( { \sqrt{3}} k_y a/2\right) }=1.
\label{nodalLine}
 \ee
 The nodal line is entirely contained in the first BZ, as shown in Fig. 2(b).  Projecting $\mathcal{H}^{(N)}$ at the lowest energy bands, the  projected Hamiltonian of the low energy excitations is   \begin{equation}
  \mathcal{H}_p(\mathbf{q}) = - [v_x(\phi) q_x + v_y(\phi) q_y ] \sigma_x +v_z(\phi) q_z \sigma_z, 
  \label{EffectiveHam}
  \end{equation} 
where $\mathbf{q}\equiv \mathbf{k}(\phi)-k_0(\phi)$, with $\mathbf{k}_0(\phi)$ defining the zero energy line, and $\sigma_x$, $\sigma_z$ the corresponding Pauli matrices acting in the projected space of the two low energy bands crossing at the Dirac loop. 

In Fig. 2(c)-(f) we show the band structure of the $\mathcal{H}$-$N$ lattices in the $k_z=0$ plane for $N= 0,\,1,\,2$ and $4$  along the path depicted in Fig. 2a. The nodal line crosses the path along both the $\Gamma-M^\prime$ and $\Gamma-R$ directions, as indicated in the plots.
As $N$ increases, the width of the unit cell in the $z$-direction increases and the corresponding height of the BZ in the $k_z$ direction decreases, but has only slight distortions in the $k_x$ and $k_y$ directions.  Therefore it is meaningful to compare plots across the \HN structures along similar paths in the $k_z=0$ plane. 

  Each additional atom in the unit cell adds an additional band to the eigenspectrum.  However, inversion symmetry forces the lowest energy bands to  cross at $E=0$, which is also the Fermi energy for the undoped system.  As $N$ increases, the optical gap between the bands  outside of the Dirac loop is greatly decreased, in particular along the $\Gamma$-$M^\prime$ direction, so that the range over which we may approximate the full Hamiltonian by a linearized model is greatly reduced, as show in Fig.\ref{fig:hhl}(c)-(f). For $N=4$  the bands quickly become quite flat along that direction away from the Dirac loop. 
  
   \begin{figure}[t]
 \centering
 \vspace{0cm}
\includegraphics[width=3.4in]{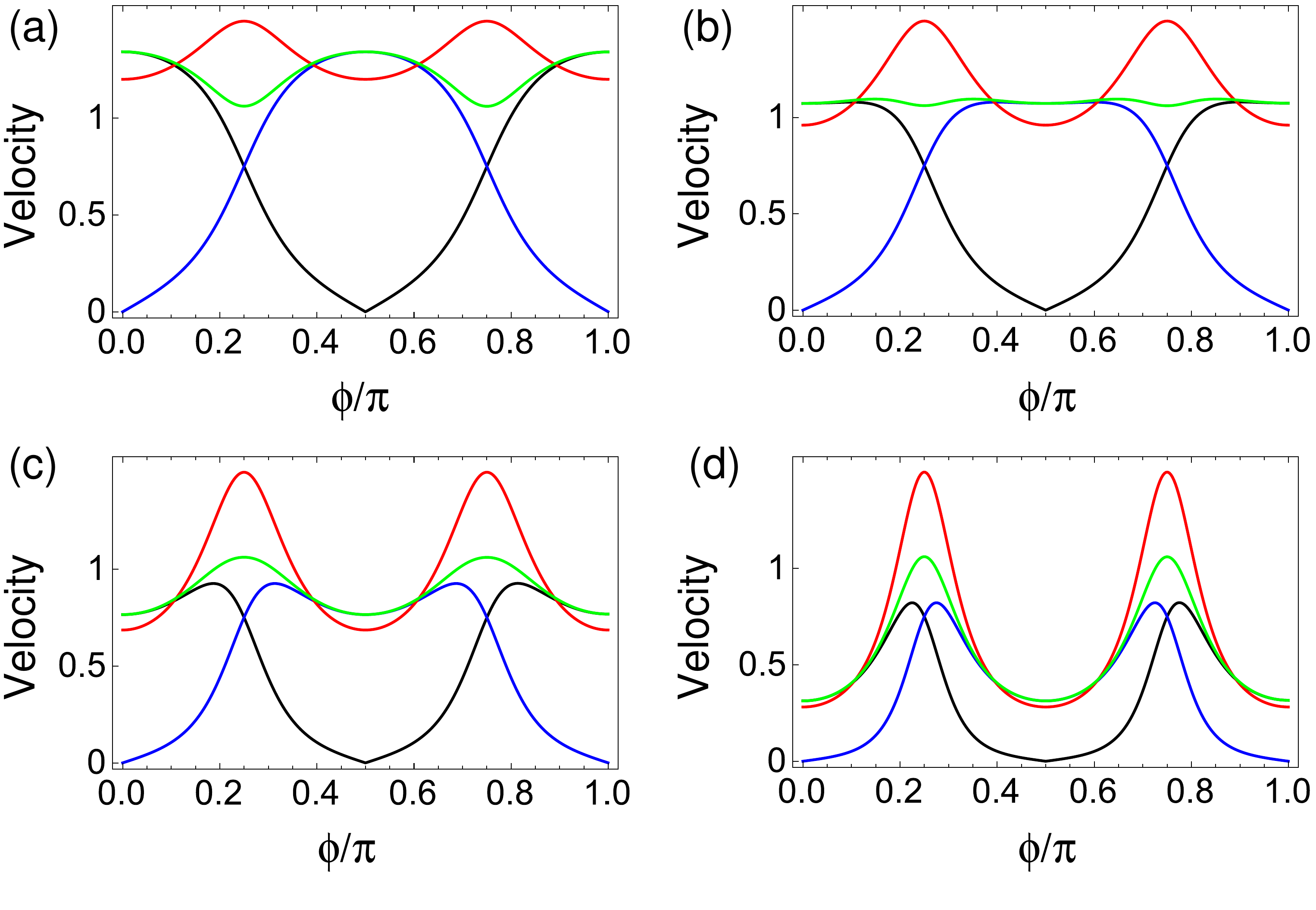} 
  \caption{{\protect\small (Color online) Velocity of quasiparticles at the Fermi energy as a function of $\phi$ in units of $t/a$, for the (a)  \H-$0$, 
    (b) \H-$1$, (c)   \H-$2$, and (d) \H-$4$ structures.  The black, blue, red and green curves are the velocities in the $v_x$, $v_y$, $v_z$ and $v_\rho$ directions repectively.     }} 
 \label{fig:velocities}
 \end{figure}

 The large $N$ limit does not trivially reduce to that of a 2D honeycomb sheet despite the presence of large parallel strips, since the six fold rotational symmetry  of stacked honeycomb lattices is never recovered.   In any event, the nodal ring persists, and the velocities in the $x$, $y$, $z$ and $\rho$ directions are plotted in Fig.\ref{fig:velocities} for $N=0,\,1,\,2$ and 4 lattices.   The ideal "isotropic" radial velocity nearly occurs with the \H-$1$ structure, with velocity asymmetries inverting from $N=0$ case, and growing stronger with $N$.   However, the 
shrinking of the gap outside the nodal ring  implies that the region over which the dispersion is linear decreases, including in the $z$ direction, which eventually becomes flat sufficiently far away from the nodal line.   

Generically, the  asymmetric $(n_x,n_y)$ structures ($n_x\neq n_y$) also have Dirac loops, whenever the onsite orbitals connecting adjacent rotated strips preserve the reflection symmetry with respect to the plane of the corresponding strip, as in the case of the $\mathcal{H}$-$N$ structures.  However, the Dirac loop is not in the $k_z=0$ plane, but is offset in the BZ in a direction that depends upon the nature of the unit cell. In Fig.\ref{fig:asymm}(d),  we show the single particle energies as a function of momentum for the path traced in momentum space in the $k_x=0$ plane, as depicted in Fig.\ref{fig:asymm}(b). The Dirac loop is not centered around the $\Gamma$ point but around $C$, at the edge of the BZ. 
All the structures  open a gap when sublattice symmetry is broken, as in the  \HN lattices.  However, the asymmetric structure differs from the \HN structure in that it lacks a screw axis symmetry.

 \begin{figure}[t]
 \centering
 \vspace{0cm}
\includegraphics[width=3.2in]{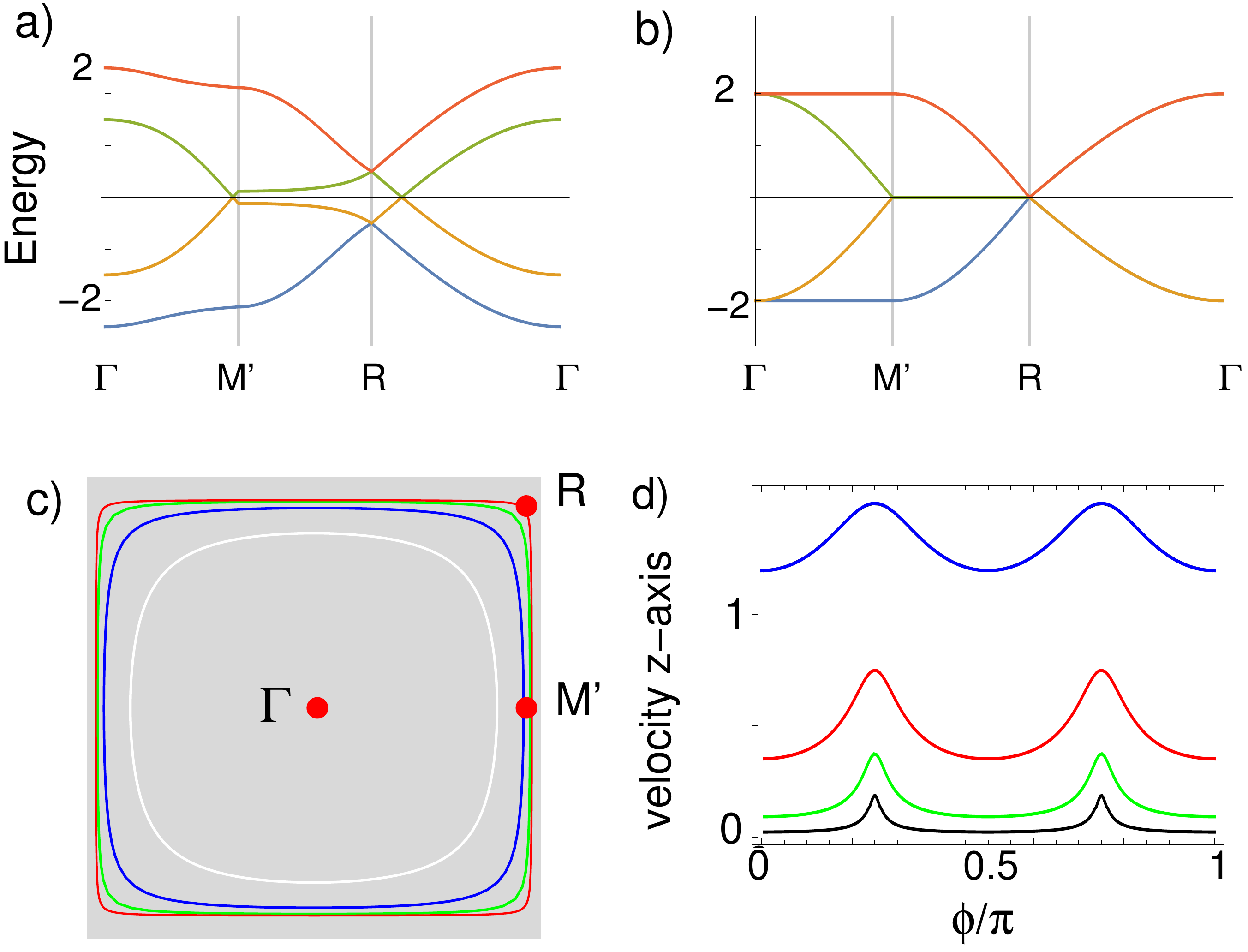}
  \caption{{\protect\small (Color online) (a) Tight bnding band structure in the $\mathcal{H}$-0 lattice for $\gamma=\frac{1}{2}$ and (b) $\gamma=0$. As $\gamma$ decreases, the zero energy line moves to the edge of the BZ, where the bands are flat. In the $\gamma\to 0$ limit, the nodal line disappears.  (c) Nodal lines in the $k_z=0$ plane as a function of $\gamma$.  White line: $\gamma=1$; blue: $\gamma=\frac{1}{2}$; green: $\gamma=\frac{1}{4}$; red: $\gamma=\frac{1}{8}$.   (d) Velocity in the $z$-direction ($v_z$) in units of $ta$ along the nodal line ($\phi$ angle) for $\gamma=1,\,\frac{1}{2},\,\frac{1}{4}$ and $\frac{1}{8}$, from top to bottom.}} 
 \label{fig:olap}
 \end{figure}

\subsection{Higher angular momentum  states}

  Conduction in 2D graphene comes from the overlap of  parallel  $p_z$ orbitals in adjacent atoms.  In the family of the hyper-honeycomb lattices there is a $\pi/2$ rotation between the $p_z$ orbitals at the edge of two adjacent perpendicular ribbons. Due to the antisymmetric nature of those orbitals, a $\varphi=\pi/2$ rotation along the screw axis leads to exact cancelation of the hopping matrix element $\gamma t=0$.  Other orbitals of higher angular momenta may lead to different, non-zero values of   $\gamma$, which parametrizes the coupling between the strips. To systematically understand this physics from the perspective of tight binding lattice models, we vary the parameter $\gamma$  in the $\mathcal{H}$-0 structure for $\gamma$ ranging from $\gamma=1$ down to $\gamma=0$.  
  
   In Fig. \ref{fig:olap} we see that as $\gamma$ decreases, the nodal ring expands to the very edge of the Brillouin zone, as illustrated in Fig. 5(a)-(c),  and the optical gap outside the nodal ring closes.  At $\gamma=0$ the ring is exactly at the edge, and at the same time, the bands crossing the nodal line become flat. In panel 5(d) we show the  velocity of the quasiparticles in the $z$ direction, along the nodal line, for different values of $\gamma$. Hence, in the $\gamma\to 0$ limit the Dirac loop is no longer present. 
 
\begin{figure}[t]
 \centering
 \vspace{0.cm}
\includegraphics[width=3.3in]{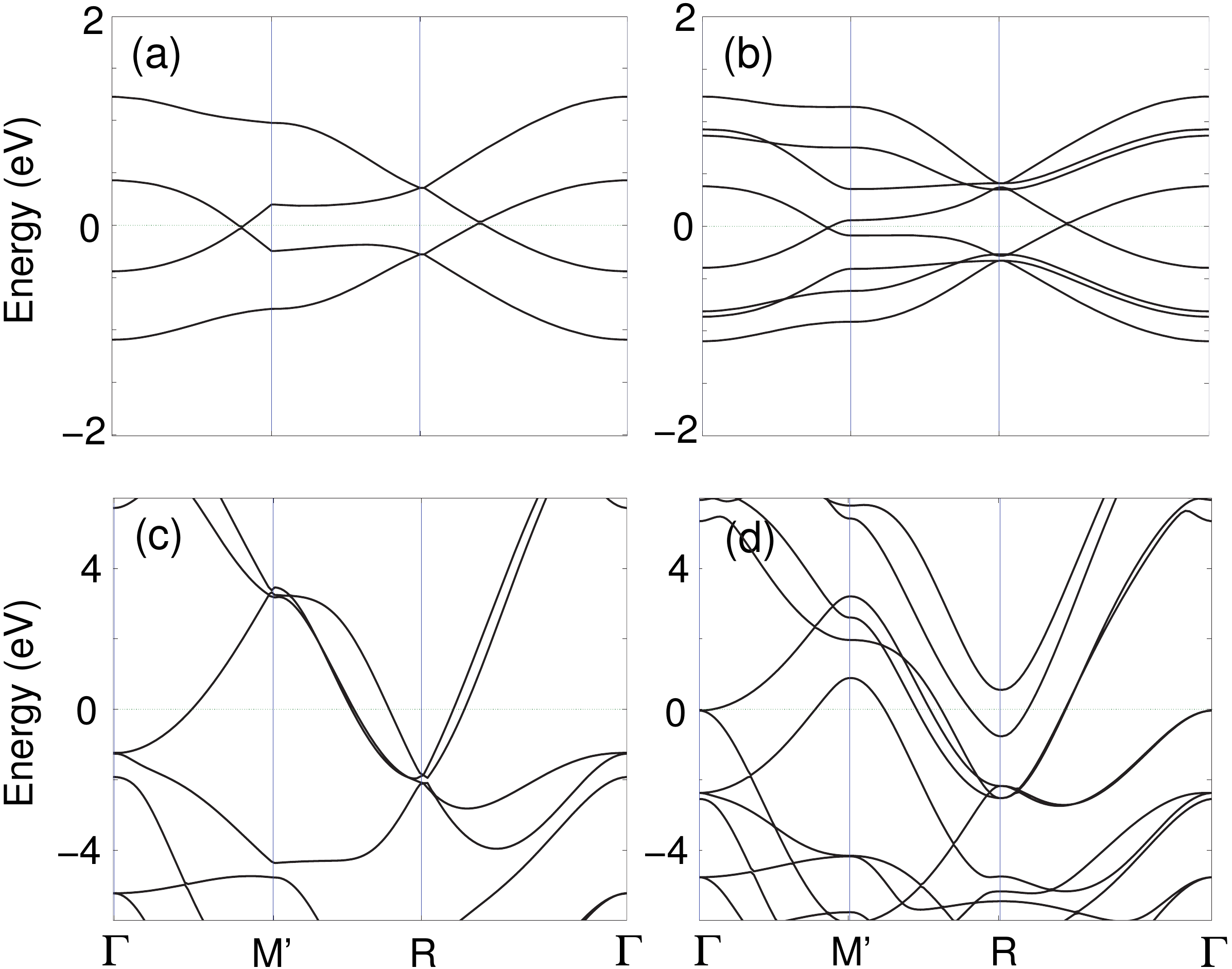}

 \caption{{\protect\small Top panels: DFT calculations for the \H-$0$ (a) and \H-$1$ (b) structures using
hydrogen-like orbitals.    In both cases the DFT calculation reproduces the structures found
in the simpler tight-binding model. Lower panels: DFT calculations for the \H-$0$ (c) and \H-$1$ (d) structures with carbon
atoms.   In both cases the suppression of inter-ribbon hopping has moved the nodal ring to the very corner of the Brillouin zone.     }} 
 \label{fig:DFTcarbon}
 \end{figure}

To verify this behavior, we conducted DFT calculations in the $\mathcal{H}$-0 and $\mathcal{H}$-1 lattices assuming both  $s$-wave and $p$-wave orbitals, as shown in the panels of Fig. \ref{fig:DFTcarbon}. The DFT results were carried out using the VASP package \cite{Kresse1}. The Perdew-Burke-Ernzerhof generalized gradient approximation exchange-correlation potential \cite{Burke} was used, and the electron-core interactions were treated in the projector augmented wave method \cite{Blochl, Kresse}.  

In panels \ref{fig:DFTcarbon}a and \ref{fig:DFTcarbon}b we used hydrogen-like atom potentials ($s$-wave) arranged in the $\mathcal{H}$-0 and  $\mathcal{H}$-1 lattices. The lattice constant was assumed to be large enough  to avoid charge polarization effects among hydrogen atoms, which could lower the point group symmetry of the crystal. The plots are in very good agreement with tight binding results. In panels  \ref{fig:DFTcarbon}c and  d, we display the DFT results for $p_z$ orbitals (carbon) for the same lattices.  While those plots also indicate the disappearance of the nodal lines, they also suggest the non-trivial role of the sigma bands in the electronic spectrum. In any case,  the crossing point is moved away from the Fermi energy and into the very corner of the Brilllouin zone. Thus,  this generic family of lattices  has nodal rings with $s$-wave states and possibly with higher angular momentum states that preserve all reflections symmetries in the unit cell. Dirac loops however cannot  be realized in simple carbon structures of those lattices, as originally conjectured \cite{DiracLoopPRL}.

\section{Screw Hyper-honeycomb Lattices}
\label{sec:screw}

 \begin{figure}[t]
 \centering
\includegraphics[width=3.0in]{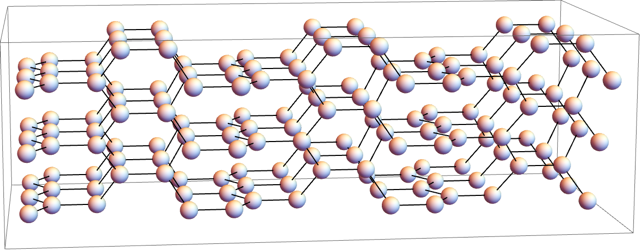}
  \caption{{\protect\small (Color online)  Screw hyper-honeycomb lattice (0,0,0). Parallel chains  of the $\mathcal{H}$-0 structure are rotated by $2\pi/3$ between the layers. The lattice has a helical  structure along the screw axis. }} 
 \end{figure}

All structures discussed up to this point have consisted  of alternating layers of parallel ``strips'' that are oriented at right angles  to one another, where the strips are either identical or different.  However, we may also create a structure in which subsequent layers are rotated by $2\pi/3$.    Such structures, the screw hyper-honeycomb lattices,  are chiral and possess a screw symmetry. The simplest example is the lattice depicted in Fig. 7, the screw hyperhoneycomb lattice, which we denote as $(0,0,0)$, where parallel chains of the $\mathcal{H}$-0 lattice are rotated by $2\pi/3$. This lattice belongs to a broader family of trigonally connected screw lattices, namely $(n_1, n_2,n_3)$, with $n_i\in \mathbb{N}$ indexing the number of honeycomb hexagons along the width of each strip.  

Those structures lack inversion and reflection symmetry along the screw axis, and hence a Dirac loop is not expected at the $k_z=0$ plane. In Fig. 8.a, we show the unit cell of the $(0,0,0)$ lattice, which has six sites. The BZ has a simple hexagonal form and has six high symmetry points, $\Gamma, K,\,M,\,A,\,H$ and $L$, shown in panel 8c. The generators of the BZ are the vectors $\mathbf{b}_1 = (4\pi/3,0,0)$, $\mathbf{b}_2 = (-2\pi/3, 2\pi/\sqrt{3},0)$, and $\mathbf{b}_3 = (0,0,4\pi/9)$, shown in the same panel. 

This lattice has a Dirac loop centered around the $A$ point, at the $k_z=4\pi/9a$ plane (see Fig. 8b). For $s$-wave states ($\gamma=1$), the Dirac loop obeys the equation
\begin{equation}
4 \cos\left( \frac{3}{2} k_x \right) \cos \left( \frac{3}{2} k_y \right) +2 \cos(\sqrt{3} k_y) = -1.
\end{equation}
This curve has six-fold rotational symmetry in the $xy$ plane.  Projection of the 6$\times$6 Hamiltonian into the two lowest energy bands results in the same low energy 2$\times$2 Hamiltonian of Eq. (\ref{EffectiveHam}).  
 \begin{figure}[t]
 \centering
 \vspace{0cm}
\includegraphics[width=3.2in]{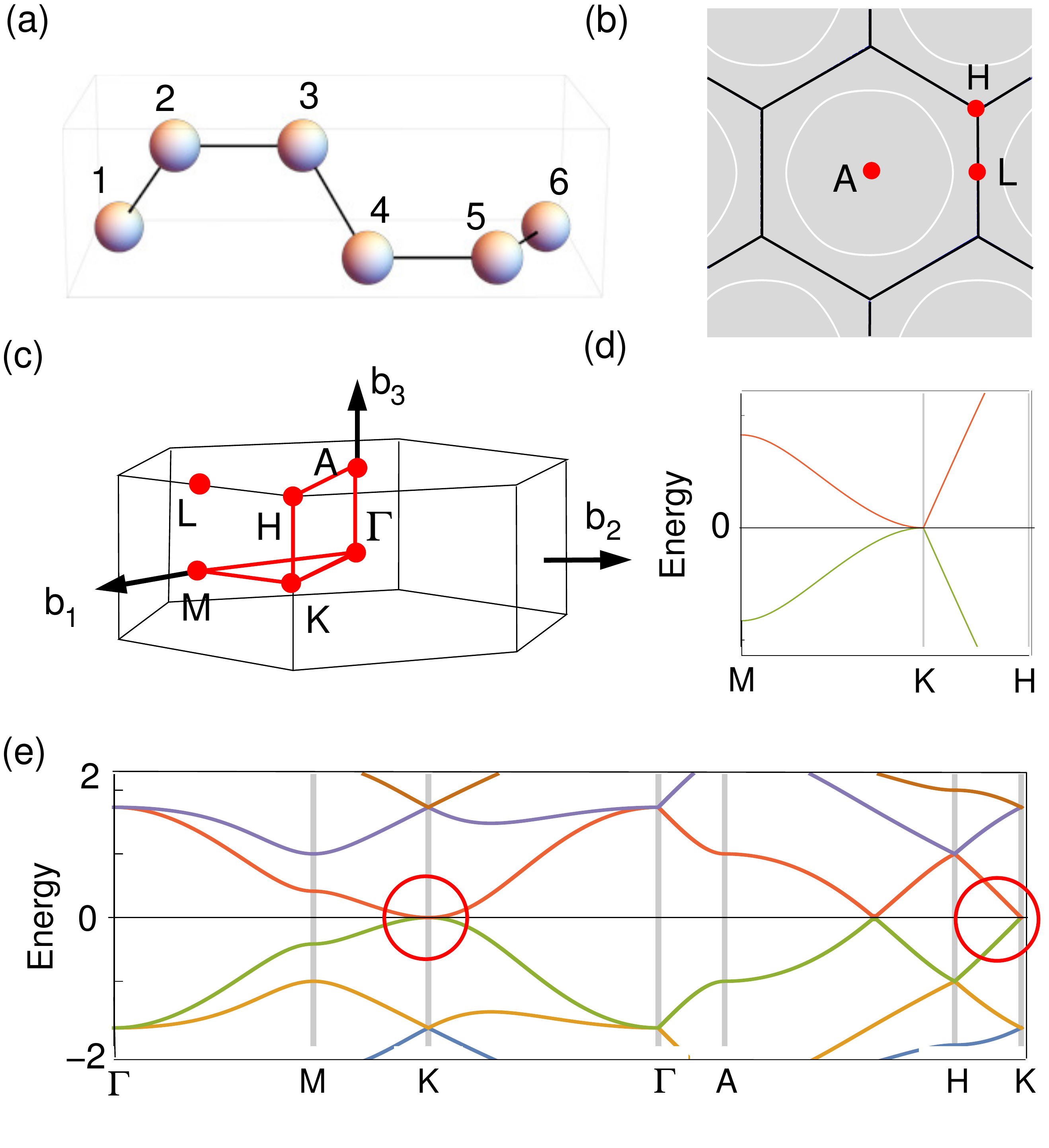}
  \caption{{\protect\small (Color online)  
  (a)  The unit cell of the screw $(0,0,0)$ structure.  (b) Slice of the BZ in the $k_z=4\pi/9a$ plane. The white circles indicate the Dirac loop for  $s$-wave states ($\gamma =1$). (c) 3D BZ of the twisted lattice, with hexagonal symmetry.    
    The higher rotational symmetry of the lattice has yielded a more circular Dirac loop.  (d) Energy dispersion around the $K$ points, indicated by the red circles in panel e. The spectrum is parabolic along the $x,y$ plane and linear in the $z$ direction.  (e) Tight-binding energy bands along the path shown in panel b  ($\gamma=1$). }} 
 \label{fig:rot3plus3}
 \end{figure}

 In Fig. 8e we show the tight-binding band structure along the path indicated in the red lines in Fig. 8(b).   The nodal line crosses the path  along the $A-H$  direction in the $k_z=4\pi/9a$ plane. In addition to the Dirac loop, the band structure also shows crossings at the $K$ points, as indicated in the red circles of panel 8(e). At those points, the spectrum of excitations is parabolic along the $x,\,y$ directions but is linear along the $z$ axis (Fig. 8d). Expanding around the two-fold degenerate $K$ points in lowest order in momentum, the 6$\times$6 Hamiltonian can be reduced to a 2$\times$2 effective Hamiltonian,
 \begin{equation}
\mathcal{H}_K(\mathbf{q}) =  -  \left [\frac{q_x^2+ q_y^2}{2m} \sigma_x + v_z q_z \sigma_y\right],
\label{9}
\end{equation} where $v_z = 3ta/2$, $m=8a^2/3t$ is the effective mass of the quasiparticles in the $k_z=0$ plane and $\mathbf{q}$ their momentum away from $K$.  This low energy Hamiltonian can be parametrized in terms of two Pauli matrices, has a chiral structure in the $zx$ and $yz$ planes and zero Chern number. Hence, those touching points are not protected by adiabatic deformations that lower the point group symmetry of the lattice, as in double Weyl points \cite{Chen}.  For $\gamma<1$, those touching points split into small Dirac loops centered around the $K$ points, as we discuss in section III.1

\begin{figure}[t]
 \centering
 \vspace{0cm}
\includegraphics[width=3.2in]{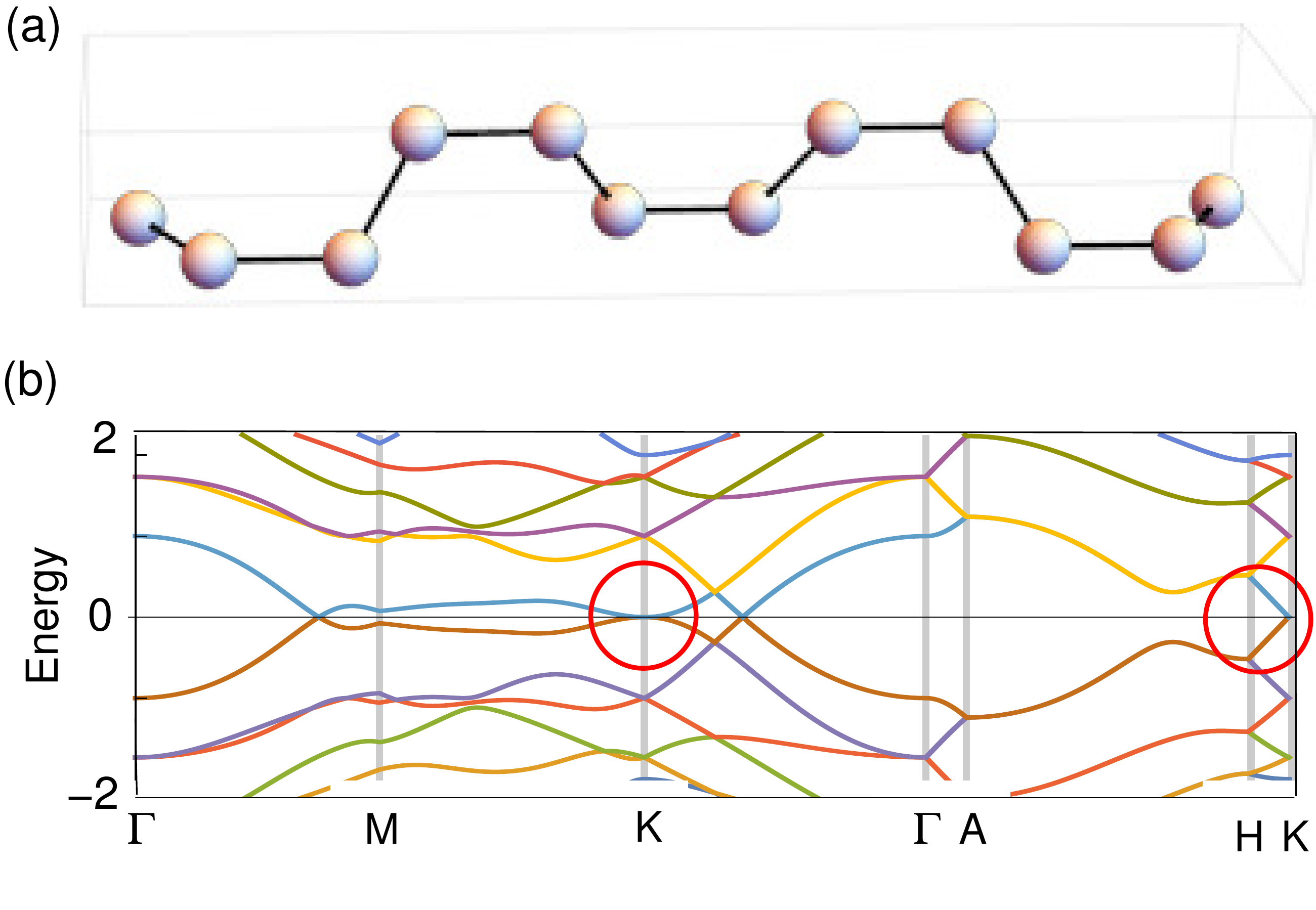}
  \caption{{\protect\small (Color online)  Generalization of the screw hyper-honeycomb lattice to one with reflection symmetry along the screw axis.  
  (a)  Unit cell with 12 sites. (b) Tight-binding energy bands along the path shown in red in Fig. 8.(c) ($\gamma=1$. }} 
 \label{fig:rot3plus3}
 \end{figure}

The screw structure of successive rotations in one direction breaks reflection symmetries and shifts the nodal line from the $k_z=0$ plane.  We may restore reflection symmetry by doubling the size of the unit cell, so that the second half consists of layers that rotate in the opposite sense to the first.  In fig.  9(a) we show the unit cell that doubles the unit cell shown in Fig. 8a along the screw axis. The extended unit cell  consists of 12 atoms sitting in a parallel chains which are rotated by $2\pi/3$ between the first three layers, and then by 
  $-2\pi/3$ between the next three layers. This lattice  generates a  similar hexagonal
crystal with a Brillouin zone plotted in Fig. 8(c).  
For $s$-wave orbitals ($\gamma=1$), the lattice has zero energy nodal lines in the $k_z=0$ plane  shown in Fig. 8(c). The equation for this loop is given by:
\begin{equation}
 4 \cos\left( {\sqrt{3}\over 2}k_y \right) + 2 \cos\left({3\over2}k_x\right)= \sqrt{6+2 \cos{3k_x}}.
 \end{equation}
 The nodal line  is centered around the $\Gamma$ point, but is not as simple as the Dirac ring expression ($\ref{nodalLine}$) in the family of the hyper-honeycomb lattice $\mathcal{H}$-$N$,  which has
 four-fold symmetry and thus are symmetric in $x$ and $y$. The  higher rotational symmetry of the lattice has yielded a more circular Dirac loop. As in the simple screw 
 $(0,0,0)$ lattice, the 12$\times$12 Hamiltonian can be projected into the same low energy 2$\times$2 Hamiltonian of Eq. (\ref{EffectiveHam}).

  \begin{figure}[t]
 \centering
 \vspace{0cm}
\includegraphics[width=3.3in]{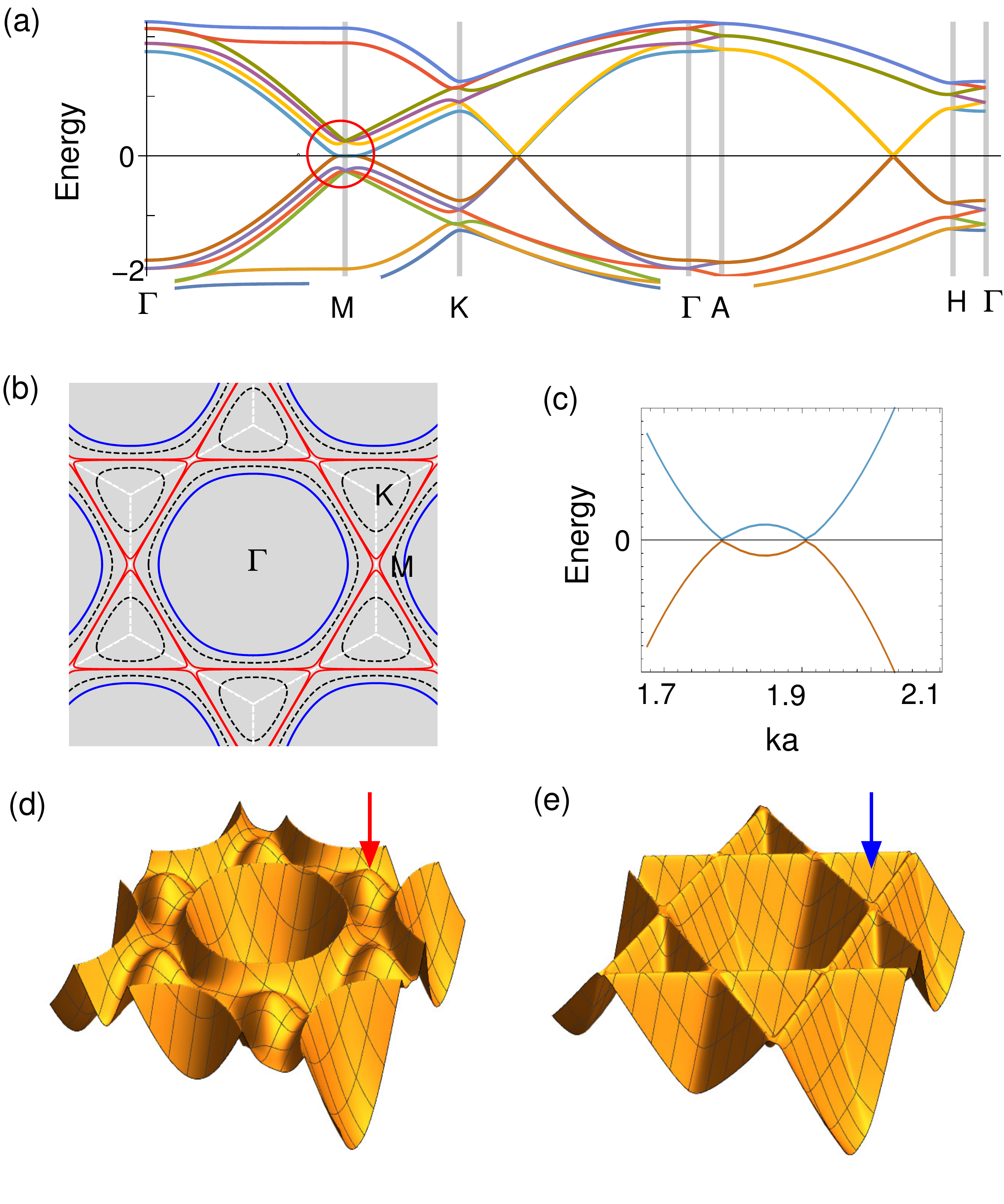}
  \caption{{\protect\small (Color online)  Tight binding band structure of the screw hyper-honeycomb lattice depicted in Fig. 9 for higher symmetry orbitals ($\gamma=1/4$) with $2\pi/3$ rotation.  
    (b) Evolution of the nodal lines with $\gamma$: Blue lines: Dirac loop for $\gamma=1$; dashed black line: $\gamma=0.75$;  and red lines: $\gamma=1/4$ lines. At $\gamma < 1$, the double Weyl points at the corner of the BZ ($K$) split into a small Dirac loop centered around $K$ and equivalent points.    (c) inset showing the merging of the the Dirac loop centered at $\Gamma$ with the one centered at $K$ along the $\Gamma-M$ line (red circle in panel a). (d) Energy surface in the $k_z=0$ plane for $\gamma=1$ in the $s$-wave case. The red arrow indicates the quadratic band touching point at $K$. (e) Energy surface in the $\gamma=1/4$ case. The blue arrow indicates the $K$ point in the center of the smaller Dirac loops.  }} 
 \label{fig:rot3plus3bs}
 \end{figure} 

\subsection{Higher angular momentum states}
  
 In panel 10(a), we show the energy spectrum for the twisted hyper-honeycomb lattice for $\gamma=\cos^2 (2\pi/3)=1/4$, which parametrizes the effect of a screw axis rotation by $2\pi/3$  between two adjacent  $p_z$ orbitals.  Unlike in the family of the hyper honeycomb lattice, which has screw axis rotations of 90$^\circ$, the hopping matrix element between two strips of atoms rotated by $2\pi/3$ is always finite, regardless the orbital symmetry.  For simplicity, we analyze the lattice with reflection symmetry in the screw axis, shown in Fig. 9(a). The analysis  for the $(0,0,0)$ structure in Fig. 8(a) is similar, except for the fact that the Dirac loop is centered around the $A$ point in the $k_z= 4\pi/9a$ plane, rather than at $k_z=0$.  
 
The evolution of the nodal lines in the twisted hyper-honeycomb lattice  is shown in panel 10(b) for different values of $\gamma$.  The blue lines represent the  $\gamma=1$ case ($s$-wave states). For $\gamma<1$, the touching points described by Eq. (\ref{9}) split into small Dirac loops centered at the $K$ point (dashed line). For $\gamma=1/4$, the Dirac loops centered at $\Gamma$ and $K$ start to merge into each other.  In panel 10(c), we zoom in  two zero energy crossings along the $\Gamma - M$ direction near the $M$ point. Each crossing belongs to a different Dirac loop.   In panels 10(d) and (e) we show the energy surfaces in the $k_z=0$ plane for $\gamma=1$ and $1/4$ respectively. The arrows indicate the position of the $K$ points at the edge of the BZ. Those panels illustrate the evolution of the bands around the $K$ points, from a parabolic dispersion in the $k_z=0$ plane to the emergence of a second small Dirac loop centered at $K$.  

The same evolution of the six touching points into small Dirac loops centered around $K$ is also present in the $(0,0,0)$ lattice, which explicitly breaks reflection symmetry. In that case the Dirac loop centered around $A$ does not directly merge with the smaller loops centered at $K$, since  both loops are located in different slices of the BZ along the $k_z$ axis.  

\section{Discussion}

The family of the hyper-honeycomb lattice with Dirac nodal lines can be experimentally realized for instance in optical lattices with fermions (such as K atoms) occupying $s$-wave states.  In optical lattices, each site can be approximated by a confining  parabolic potential. $s$-wave states can be physically constructed in the situation where the lowest energy Bloch band is populated. 

Very recently, honeycomb crystals have been artificially simulated with fermionic optical lattices using three interfering laser beams with appropriate phases, amplitudes and detuning \cite{Tilman}. Those experiments measured the energy of the Dirac quasiparticles and their corresponding Berry phases. Because of the tunability of the system, the Dirac points that result can be merged and manipulated.  Other experimental results \cite{Tilman2} include the realization of the Haldane model, where the bands acquire a Chern number and become topological. In theory, the generalization of the three laser beam geometry to the hyper-honeycomb lattice $\mathcal{H}$-0 may provide a simple physical realization of this system. The population of the lowest energy band with fermionic atoms constitutes a promising path towards the experimental observation of a Dirac loop in this geometry.  

We showed that nodal lines can be realized in $s$ and higher angular momentum states in the screw hyper-honeycomb structure with $2\pi/3$ rotations. Higher angular momentum states such as $p$-wave could be in principle realized in the excited Bloch bands of optical lattices  \cite{Kock, Wirth}.  Those bands are metastable and have a lifetime in the ms range, shorter than the typical intraband relaxation time scales.  Those states have been physically realized in bosonic bipartite lattices  engineered with a shallow and a deep potential wells  \cite{Kock, Wirth}. This construction breaks sublattice symmetry, and cannot be used for observing nodal lines. In the screwed hyper-honeycomb case, the two sublattices must have potentials of equal depth.  

In summary, we have systematically described the nodal lines that appear in different families of 3D crystals with planar trigonally connected sites.  In the family of the hyper-honeycomb lattice, we showed that Dirac lines are expected in the presence of on site orbitals that preserve the reflection symmetry of the planes set by the trigonal links. We found that screwed hyper-honeycomb structures also present Dirac loops regardless the symmetry of the on site states. Those crystals also present a touching point at the corners of the BZ with chiral 3D quasiparticles, but zero Chern number. Those lattices are just a small subset of a vast number of possibilities of tight-binding lattices having nodal quasiparticles. We have proposed optical lattice realizations of those structures.

\section{ Acknowledgements} We thank  I.~Spielmann,  A.~Schwettman and T.~Esslinger for illuminating discussions. K. M. was supported by NSF grant DMR-1310407. B. U. acknowledges NSF CAREER grant DMR-1352604 for support.  The DFT calculations were performed at the OU Supercomputing Center for Education and Research at the University of Oklahoma.

\end{document}